# Towards decoding the relationship between domain structure and functionality in ferroelectrics via hidden latent variables

Sergei V. Kalinin,[1,*] Kyle Kelley,[1] Rama K. Vasudevan,[1] and Maxim Ziatdinov[1,2,*]

[1] The Center for Nanophase Materials Sciences, Oak Ridge National Laboratory, Oak Ridge, TN 37831

[2] The Computational Sciences and Engineering Division, Oak Ridge National Laboratory, Oak Ridge, TN 37831

**Abstract**

Polarization switching mechanisms in ferroelectric materials are fundamentally linked to local domain structure and presence of the structural defects, which both can act as nucleation and pinning centers and create local electrostatic and mechanical depolarization fields affecting wall dynamics. However, the general correlative mechanisms between domain structure and polarization dynamics are only weakly explored, precluding insight into the associated physical mechanisms. Here, the correlation between local domain structures and switching behavior in ferroelectric materials is explored using the convolutional encoder-decoder networks, enabling image to spectral (*im2spec*) and spectral to image (*spec2im*) translations via encoding latent variables. The latter reflect the assumption that the relationship between domain structure and polarization switching is parsimonious, i.e. is based upon a small number of local mechanisms. The analysis of latent variables distributions and their real space representations provides insight into the predictability of the local switching behavior, and hence associated physical mechanisms. We further pose that the regions where these correlative relationships are violated, i.e. predictability of the polarization dynamics from domain structure is reduced, represent the obvious target for detailed studies, e.g. in the context of automated experiments. This approach provides a workflow to establish the presence of correlation between local spectral responses and local structure, and can be universally applied to spectral imaging techniques such as PFM, scanning tunneling microscopy (STM) and spectroscopy, and electron energy loss spectroscopy (EELS) in scanning transmission electron microscopy (STEM).

**Introduction**

For almost a century, ferroelectrics remain one of the most fascinating classes of materials both due to the broad variety of application-relevant physical behaviors they demonstrate[1-3], and the set of fundamental scientific issues they exemplify.[1, 4] The presence of strong electromechanical responses, optoelectronic couplings, and particularly switchable polarization enable applications of ferroelectrics in transducers and microelectromechanical systems (MEMS), optoelectronic devices, and ferroelectric memories.[5, 6] At the same time, the unique physics of ferroelectrics driven by the presence of switchable polarization and strain, and their interaction with the non-local electrostatic and mechanical depolarization fields have been a focus of extensive research for several decades. In single crystals and thin films, these non-local mechanisms result in the emergence and evolution of complex domain structures and can give rise to a rich variety of topological defects with unique electronic properties.[7, 8] Further, in the presence of structural disorder or competing interactions, these defects can give rise to frustrated ground states with giant functional responses and a broad distribution of relaxation times, as exemplified by ferroelectric relaxors[9-13] and morphotropic phase boundary materials.[14-20]

On the macroscopic level, polycrystalline and disordered ferroelectrics have been extensively studied via frequency and time-resolved spectroscopies, revealing a broad range of relaxation timescales associated with the responses of individual topological defects, their interaction with structural defects, and collective effects induced by interactions mediated by long-range electrostatic and strain fields.[4, 6, 21] Further, scattering studies via X-ray and neutron diffraction allow for the identification of globally averaged aspects of the polarization order parameter including motion of the ferroelastic domain walls and polarization rotation.[22-24] As such, these considerations necessitate exploring the local relationship between the domain structure and functionalities.

The emergence of Piezoresponse Force Microscopy (PFM) in the 1990s[25-28] provided a powerful tool for real space visualization of ferroelectric domain structure and domain manipulation, with multiple examples of domain structures in ceramics,[29] polycrystalline[30] and epitaxial films,[31, 32] and disordered ferroelectrics[33, 34] being reported. The subsequent development of Piezoresponse Force Spectroscopy (PFS)[35] enabled direct measurements of the local electromechanical hysteresis loops, opening the pathway for exploring the correlations between local microstructural elements and polarization switching mechanisms. Later development of switching spectroscopy PFM[36] and its band-excitation version[37] have enabled hyperspectral imaging in PFM, where hysteresis loops can be acquired over dense rectangular grids, thus visualizing switching behaviors. Multiple studies of ferroelectric switching behavior over single isolated defects,[38] grain boundaries,[39] ferroelectric and ferroelastic domains walls,[40] and collective phenomena in ferroelectrics[41] have been demonstrated using these techniques, providing deep insights into local polarization switching mechanisms.

However, the proliferation of the PFM and hyperspectral PFS modes, including multidimensional time- and voltage spectroscopies brings forth the challenge of systematic analysis of associated mechanisms. In the spectral domain, the use of multivariate statistical methods such as principal component analysis (PCA)[42] and non-negative matrix factorization (NMF) have allowed dimensionality reduction of the D-dimensional data sets to yield a linear combination of 2D spatial loading maps and spectral components.[43] Here, the components define the specific time- and voltage behaviors, whereas the loading maps show its spatial variability. The spatial behavior of the contrast in loading maps or its comparison with the PFM images allows qualitative conclusions about underpinning mechanisms to be derived. The advantage of this approach is that the dimensionality reduction/linear unmixing tools are both highly robust and are now readily available. However, these tools are linear (which is not always the case for underpinning physics), generally ignore the correlations in the image plane (i.e. details of domain structure), and are invariant towards the redistribution of spatial pixels. This in turn precludes elucidation of specific behaviors and mechanisms unless they are clearly discernable, or specific hypothesis can be formulated and tested. The latter approach is, however, severely limited by the general dearth of quantitative models necessary for hypothesis testing. Alternatively, methods such as tensor factorization[44] can be used. Similarly, while these approaches have been used as a basis for recognition imaging,[45] physics-based unmixings,[46] or model comparison using (shallow) neural networks,[47] the analysis was performed on a single-pixel basis.

An alternative approach for the analysis of switching behaviors can be based on the domain patterns and their evolution with time under bias, thermal, or pressure stimuli, i.e. feature extraction in the spatial domain. Here, chosen physical descriptors such as distance to the ferroelectric or ferroelastic wall or local curvature are selected and the correlations between the physical descriptors and switching behavior are explored.[48] The drawback of such approach is the large number of possible mechanisms and hence possible descriptors, giving rise to spurious correlations. For completeness, it should be noted that the descriptors can be derived in an unsupervised fashion through specific machine learning algorithms, ranging from the linear unmixing to the autoencoders and variational autoencoders.[49] However, if identified, the next problem will be correlation between these descriptors and the chosen functionalities.

As such, here we introduce an approach for exploring the correlations between local domain structures and switching behavior using convolutional encoder-decoder networks, enabling image to spectral (*im2spec*) and spectral to image (*spec2im*) translations. This approach is based on the premise that (i) domain structures and polarization dynamics in ferroelectrics are related, and (ii) this relationship is ultimately parsimonious, i.e. builds upon a small number of (unknown) active local mechanisms. The encoder-decoder network structure with its latent representation bottleneck naturally comports to these assumptions. We note that while (to our knowledge) previously unexplored in the context of physical sciences, the neural networks-based transformation between different signal types were studied in the context of image-to-text[50] and video-to-speech[51] conversion. Here, in the *im2spec* approach, the sub-images representing local domain structure are compressed via classical convolutional architecture to a small number of latent variables, and the latter are deconvoluted to yield the spectra. In the *spec2im* approach, the process is reversed. The analysis of latent variables distributions and their real space representations provide insight into the associated mechanisms. Similarly, the regions where the correlative relationships fail are potentially associated with new (i.e. not represented in in the majority of spatial locations) physical phenomena, and hence offer a natural target for focused studies. Both the prediction errors and latent variables can further offer a target function for the automated experiment.

**Experimental**

As a model system, we have chosen a ferroelectric $PbTiO_3$ thin film, in the tetragonal phase. The film was grown on a $SrRuO_3$ bottom electrode on a $KTaO_3$ substrate by chemical vapor deposition and possess a dense polydomain structure as previously reported.[52] The large density of domain walls as well as the hierarchical structure lend themselves well to the studied approach. The PFM imaging and spectroscopy were performed with Pt/Ir-coated tips (k~1N/m) on a commercial microscope (Cypher, Asylum/Oxford Instruments) at room temperature with in-house developed LabVIEW code to acquire the band excitation piezoforce spectroscopy data using National Instruments hardware. The DC voltage was ramped from –12.0 V to +12.0V applied to the tip with the piezoresponse measured using the band-excitation approach with a 1V AC signal. This yields spectra in the form of real and imaginary components that are fit to the simple harmonic oscillator function to return the amplitude, phase, Q-factor and resonant frequency for each position and spectroscopic value (in this case, each DC voltage value). All piezoresponse values reported here were taken in the 'off-field' state, i.e., taken immediately after the DC voltage applied was removed. After accounting for an instrument-specific phase offset, the amplitude and phase could be combined to yield the piezoresponse, i.e. $Acos(\phi)$ where $A$ is the amplitude and $\phi$ is the associated phase of the response, yielding a spectroscopic dataset.

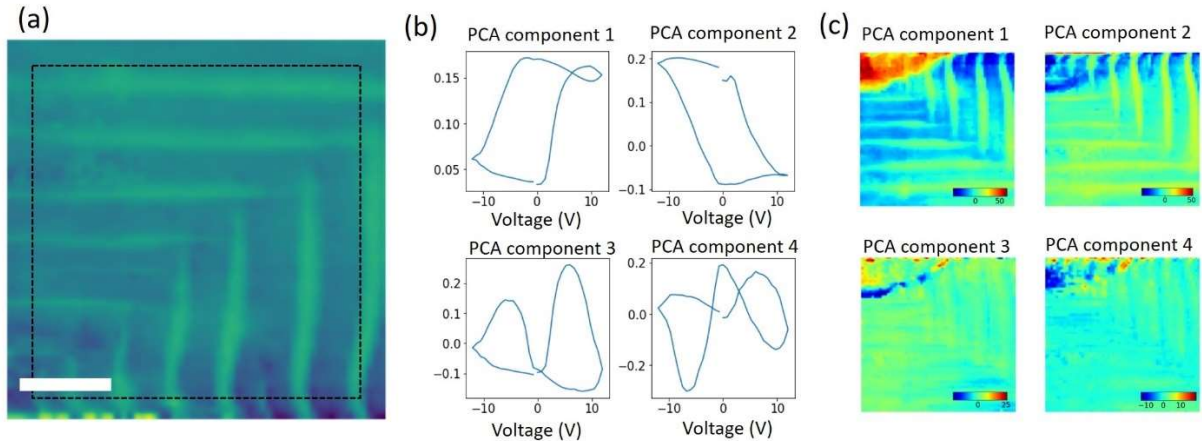

**Figure 1.** (a) The typical domain structure of the PTO sample where dark regions are in-plane $a$ domains, and bright regions are out of plane $c$ domains. Scale bar is 200 nm. The pixel coordinates within the dashed square are used to create a training set for *im2spec* and *spec2im* models (parts of training sub-images can extend outside the square). (b) First four principal components of the hysteresis loops with (c) corresponding loading maps. The additional example of analyses for the region with more complex domain structure is shown in Supplementary materials.

PFM spectroscopy was performed on a 60x60 grid of points in a region of size 750 nm x 750 nm. The spatial spectroscopy grid corresponding to the zero DC voltage pulse, is plotted in Fig. 1(a) and exhibits classical crosshatch pattern of the in-plane *a* and out of plane *c* domains. Generally, the sample shows a good switching behavior, with the hysteresis loops open and saturated over *c*-domain regions.

**Results and Discussion**

The hyperspectral PFS dataset was analyzed using Principal Component Analysis (PCA),[53] with corresponding components and loading maps shown in Fig. 1 (b-c). Note, here the PCA analysis is performed without normalizing the data since the vertical shifts of the hysteresis loop and their absolute magnitude contain the information on ferroelectric behavior (polarization imprint and switchable polarization, respectively). Although the PCA components are defined only in an information theory sense, some conclusions can be made based on the qualitative analysis. Here, the first component shows the effective loop opening, and can roughly be identified with switchable polarization. The second component represents both the loop broadening for negative biases and opening, whereas the third component yields the information on the loop shift and the fourth component yields the information on the loop width. Similar interpretations can in principle be proffered for the first ~5-6 components, while further components adopt complex shapes describing the variability of the switching behaviors in the vicinity of coercive bias. The corresponding spatial maps contain localized features associated with progressively finer details of the domain structure, suggesting the presence of a rich spectrum of local physical mechanisms.

The systematic analysis of these behaviors necessitates, as a first step, establishing the correlation between the characteristic features in the spectral and spatial domains, e.g. hysteresis loop parameters such as imprint, offset, width and/or PCA components describing the loop shape, and the domain structure descriptors. The visual examination of the components and loading maps of the PCA decomposition suggests that the spatial and spectral features can be detected even in the high-order components, up to 20-30. This conclusion can be verified e.g. by the spatial correlation function analysis, as proposed by Belianinov et al.[54] For the spatial features, some insight into possible domain structures can be obtained by methods such as linear on non-linear unmixings of image patches or their Fourier or Radon transforms. However, the simple analysis by e.g. the Gaussian mixture model[55] (GMM) as illustrated in the accompanying notebook demonstrates that the number of relevant spatial features is also very large. These descriptors are non-parsimonious, which severely impinges development of any quantitative descriptions upon it. Correspondingly, direct correlation analysis between spectral and spatial features, especially given the limited statistics of the available data sets, is unlikely to lead to fundamental physical insights.

Here, we propose to explore the relationship between the local domain structures and switching behavior using the encoder-decoder networks. In the *im2spec* approach, the sub-images representing a local domain structure are "compressed" via a convolutional neural network architecture to a small number of latent variables, and the latter are "deconvoluted" to yield the spectra. In the *spec2im* approach, the process is reversed. In both cases, the data flows through the latent variable bottleneck that compresses the data stream and selects the relevant representations of the data, while rejecting the noise and spurious features.[56] Note that in the context of classical autoencoders (both conventional[57-59] and variational[60-62]), this approach is used to disentangle the representations of the data, e.g. identify the emotional expressions in facial data bases or writing style in hand-written digit data bases.[63, 64] Recently, the autoencoder approach has been also used for the analysis of ferroelectric switching loops.[65] The capability of (variational) autoencoders to discover relevant trends in data have attracted recent attention of the theory community, but in all of these cases the autoencoder inputs and outputs are identical. Here, we explore the encoder-decoder architectures in the context of structure-property relationships, that is, *structure → latent space → property* and *property → latent space → structure* predictions, and analyze the latent space of the experimental data.

The networks are implemented using the PyTorch deep leaning library[66] and are available from the accompanying Colab notebook. The architecture of the *im2spec* model is straightforward. It consists of two parts: an encoder for embedding input images into a latent vector and a decoder for generating one-dimensional signals from the embedded features. The encoder part consists of three back-to-back two-dimensional convolutional layers, whereas the decoder part represents a pyramid of one-dimensional dilated convolutions[67] with dilation rates of 1, 2, 3, and 4 (Fig. 2). The dilated convolutions showed superior performance compared to regular convolutional blocks in the decoder. Each convolutional layer in both encoder and decoder parts is formed by running 64 filters ("kernels") of size 3 on its input and is activated by a leaky rectified linear unit with a negative slope of 0.1. The batch normalization is applied after each activation to avoid overfitting and improve generalization to the new data. The "bottleneck" layer for representing the latent vector is a fully connected layer with the number of neurons equal to the specified number of latent dimensions.

For *spec2im*, the general structure of the network remains largely the same, except that the encoder now consists of one-dimensional convolutional layers, while the convolutional layers in the decoder are two dimensional. In addition, we found that increasing the number of convolutional layers in the encoder from 3 to 4 and adding 2 up-sampling layers (consisting of regular convolution followed by *x2* up-sampling operation) before the dilated block in the decoder improves the overall accuracy (the average error becomes comparable to that of the *im2spec*). The networks were trained using *Adam* optimizer[68] with a mean square error (*mse*) loss function and a learning rate of 0.001. The size of mini-batches was set to 64. We note that in principle, the loss function can be further engineered to focus on specific aspects of the images or spectra based on known physical characteristics, e.g. optimize the reconstruction of the nucleation biases or incipient antiferroelectric behaviors defined as a non-monotonic slope in the vicinity of zero bias. This approach will be equivalent to a priori feature engineering based on known physics. However, we leave the detailed exploration of this possibility to future studies.

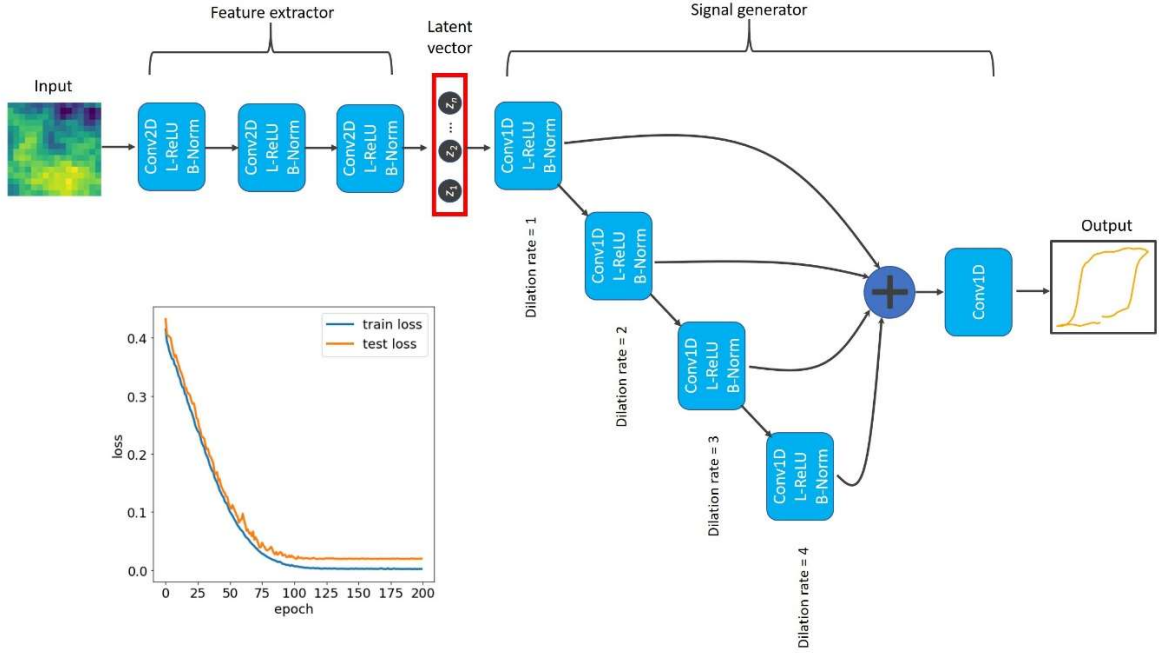

**Figure 2.** Schematics of the *im2spec* network and its characteristic training behavior. The network takes images as input and produces spectra. It can be separated in two parts: encoder for feature extraction and decoder for signal generation. The Conv2D and Conv1D are two-dimensional and one-dimensional convolutional layers, respectively, L-ReLU is leaky rectified linear unit (negative slope is 0.1) and B-Norm is batch normalization. The *spec2im* network (not shown but available from the accompanying notebook) shares a similar architecture with Conv2D and Conv1D blocks swapped; in addition, there is one more convolutional layer in the encoder and two (convolution + up-sampling) blocks in the decoder before the dilated convolutions.

The data was preprocessed in the following way. The original datasets were of the size 60x60x64 and 50x50x64, corresponding to ($x$, $y$, $V$) dimensions. Small image patches of size (8, 8) and (16, 16) centered on a chosen pixel ($x_i$, $y_i$) were taken from this dataset, corresponding to the spatial map at $V = 0$ V (edge pixels, where the image patch would fall beyond the existing dataset frame, were ignored). The full hysteresis loop spectra at the chosen pixel location ($x_i$, $y_i$) were also extracted. This process was repeated to form the data on which to perform the encoding-decoding process. This datasets were split into a training dataset and testing dataset, with a split of roughly 0.75 : 0.25 and 0.80:0.20. To avoid the spurious results associated with the overlap between the training and testing data selected for adjacent image pixels, the testing was performed with the network trained on the compact subset of the image and applied to remaining part of the image, which is equivalent to applying a model trained on data from one scan area to make a prediction for data from a different scan area of the same sample acquired with the same experimental parameters.

Insights into the *im2spec* network behavior and the quality of prediction can be obtained from the training/testing curves, the spatial map of the prediction errors as shown in Fig. 3, as well as the inspection of the individual predictions on the testing set (Fig. 4a-d). During the training, the network demonstrates good convergence with the stable behavior both for training and test sets, suggesting the existence of strong correlative relationship between the local images and hysteresis loops. The error map for full datasets in Fig. 3 (a) is relatively structureless and do not show similarity to the original domain structure or the PCA loading maps, suggesting that the *im2spec* network captures the relationship between the domain structure and the local hysteresis behavior well on average. Note, the regions with strong deviations from the network prediction can be considered as locations where novel physical behaviors can emerge, as was proposed earlier.[69]

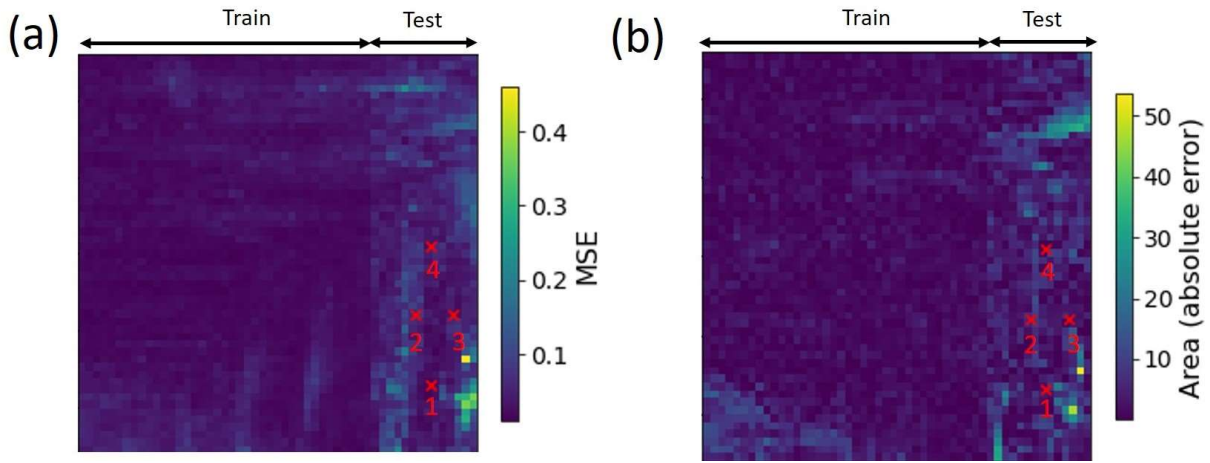

**Figure 3.** Spatial error maps for the entire dataset for *im2spec* model. (a) Mean squared error. (b) Absolute error in loop area. Note that here the test is performed at the r.h.s. part of the image, obviating the potential overlaps between the domain structures encoded in sub-images centered at adjacent locations as would be the case for classical testing.

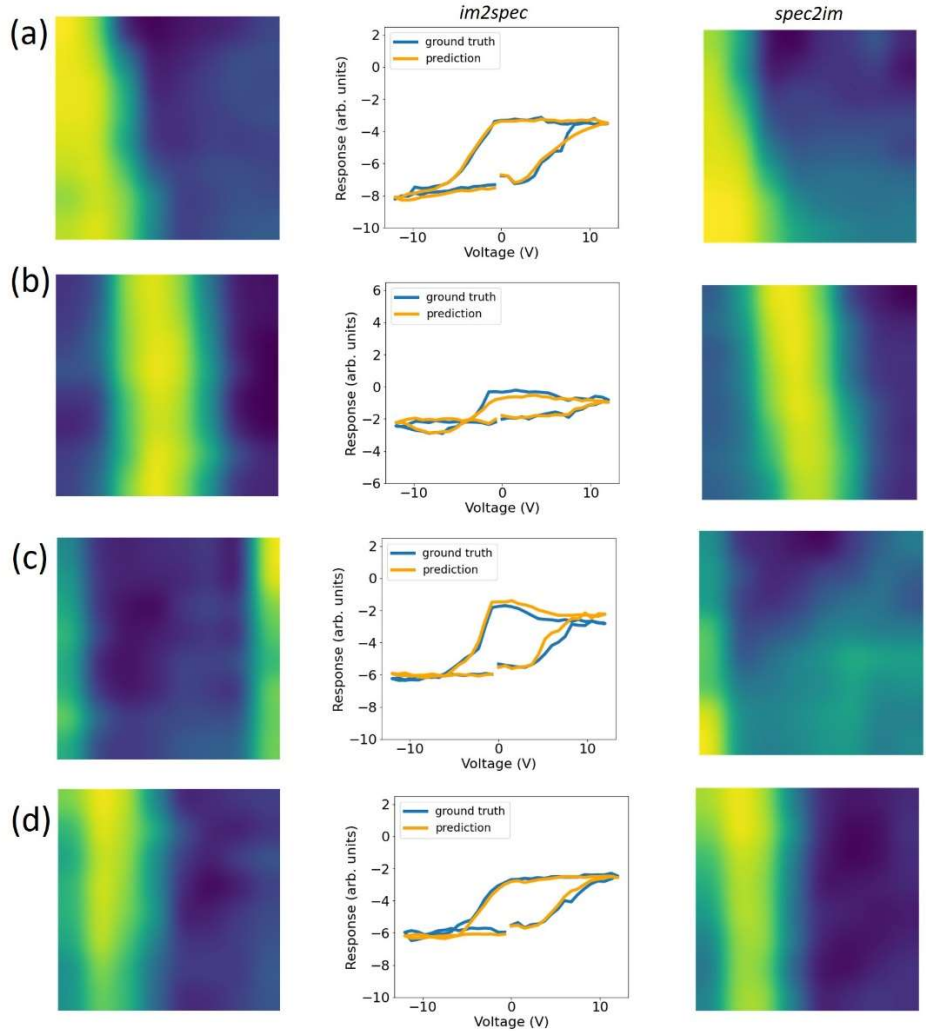

**Figure 4.** (a-d) Example of *im2spec* and *spec2im* predictions made for points 1-4 in Fig. 3. For *im2spec*, the image data in the first column are the inputs (a, c) and spectral curves in the second column are the neural network's prediction. The third column shows a "reverse" predictions of images from spectra via *spec2im* for the same points. The accompanying notebook allows making and inspecting predictions for all the inputs in the dataset.

The comparison of the predicted and ground truth hysteresis loops is shown in Fig. 4 a-d. Here, the left column shows the chosen sub-images representing easily recognizable elements of the domain structure, namely the domain walls at different separation from the location at which the hysteresis loop is acquired (image center). The central column compares the predicted hysteresis loops to the experimentally measured ones. Note that the hysteresis loop behavior is strongly dependent on location, showing strong effect of domain structure on switching. However, this variability is well captured by the *im2spec* prediction. Note that significantly more complex example of domain structure shown in Figure 5 also shows the high quality of *im2spec* prediction.

The simplicity of chosen domain region allows for a straightforward explanation of observed behaviors, following the mechanisms explored previously by Ganpule *et al*.[70] Here, the switching in the center of the *c* domain gives rise to a classical well-behaved hysteresis loop, Fig. 4 a, c. In the in-plane domain region, the switching is impeded and can be associated with the reversible bias-induced reconfiguration of the in-plane *a* domain geometry, giving rise to weak anomalous hysteresis with the characteristic "noose" shape (Fig. 4 (b)). Finally, the switching behavior further varies within the *c* domains, which is dependent on the proximity of the *a-c* domain walls due to the intrinsic asymmetry of the latter. In this case, if the wall that is crystallographically constrained to (110) plane and hence is 45º tilted under the AFM tip, switching is impeded since the domain needs to cross the wall. In contrast, if the wall does not pass under the AFM tip, normal 180º switched domain forms, which explains the difference between Fig. 4a, 4c, and 4d.

Finally, the right column in Fig. 4 shows the reverse prediction via the *spec2im* approach. Here, the network was trained using the hysteresis loops as features and domain images as targets. Interestingly, the quality of prediction in this case is also high, suggesting the presence of strong bijective correlation between domain structures and polarization dynamics.

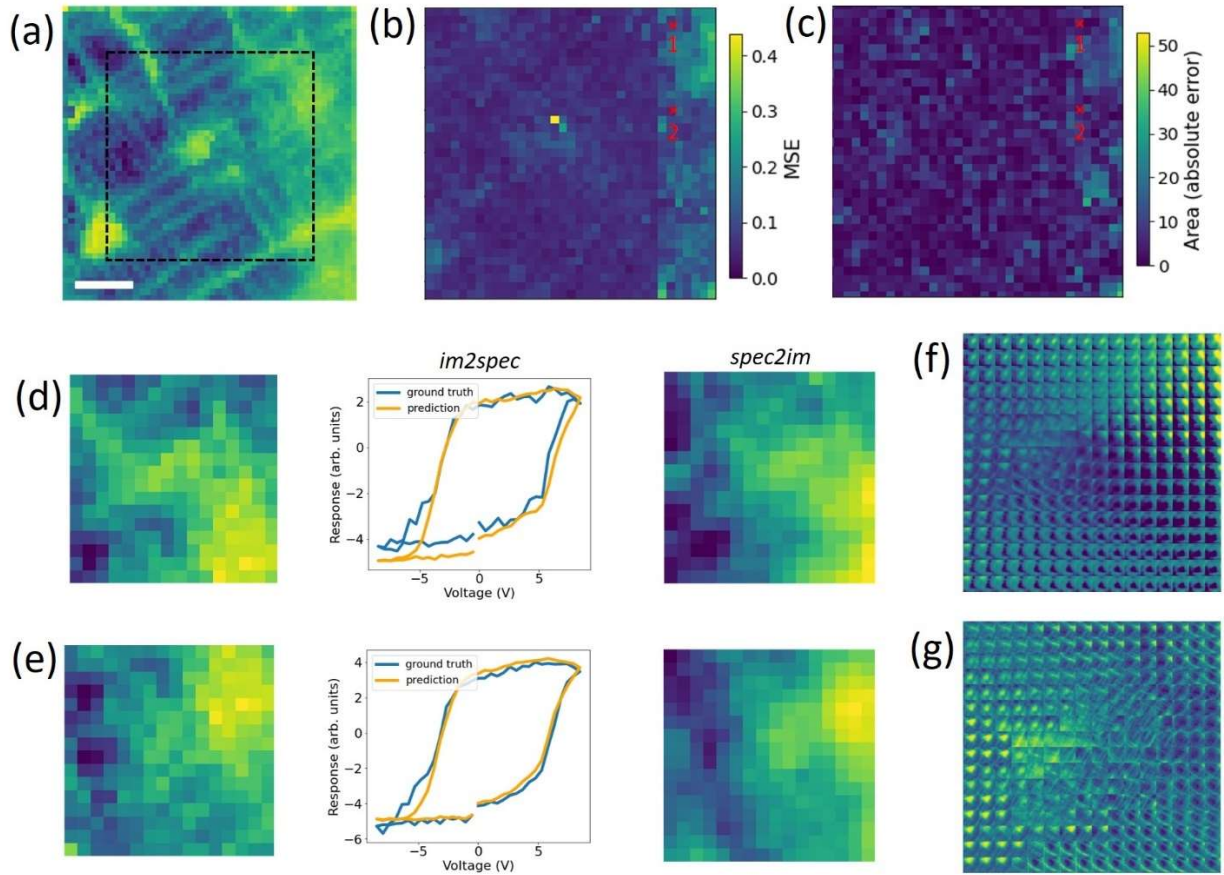

**Figure 5.** Application of *im2spec* and *spec2im* models to a different dataset with a more complex structure. (a) The domain structure at $V = 0$ V. The scale bar is 400 nm. The coordinates within the dashed square are used to create a training set (parts of training sub-images can extend outside the square). (b, c) Spatial error maps for the entire set: mean squared error (b) and absolute error in loop area (c) for every extracted subimage (d, e) Examples of *im2spec* and *spec2im* predictions made for points 1, 2 in (b, c). For *im2spec*, the image data in the first column are the inputs and spectral curves in the second column are the neural network's prediction. The third column shows predictions of *spec2im* for the same points where spectral data from the second column are the inputs and the images are the *spec2im* predictions. The accompanying notebook allows making and inspecting predictions for all the inputs in the dataset. (f) Learned 2D manifold with different domain behaviors reconstructed from latent variables by the *spec2im* decoder. (g) "inverse" reconstruction of domain behaviors with *im2spec* model encoder.

The analysis of the relationship between the local domain structure and hysteresis behavior is illustrated in Fig. 6. Here, we explore the distribution of the experimental data in the latent space of the system, i.e. reduced variables that connect the encoder and decoder parts of the network. Shown in Fig. 6 (a) is the structure of the latent space for the *im2spec* network for 2 latent variables. The distribution is generally broad and does not have clearly identifiable features. Figure 6 (b) and 6 (c) show projection of the first 2 latent variables onto the real space. One can clearly see that the first latent variable is associated with *a* domains, whereas the second one is associated with *c* domains.

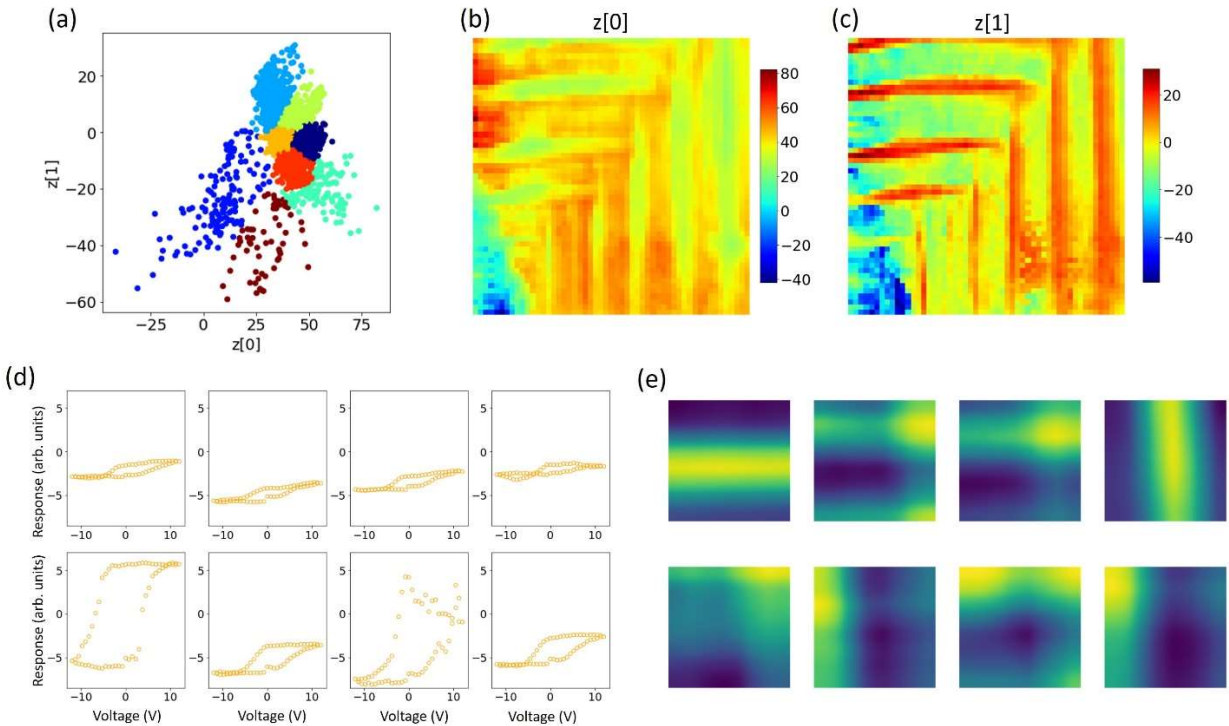

**Figure 6.** (a) Latent space for the *im2spec* network for 2 latent variables. Different colors correspond to different clusters in Gaussian mixture model. (b, c) Value of the first (b) and second (c) latent variable for each data point (train+test) projected to the image space. These variables serve as hidden parameters enabling the connection between local domain structures and hysteresis

loops. (d) Characteristic hysteresis loops derived by applying *im2spec* decoder to the Gaussian mixture model-computed centroids in latent space. (e) The corresponding domain structure elements.

To obtain further insight into the relevant domain and spectral behaviors, the latent space is separated using a GMM. We note that metrics-based algorithms such as *k*-means clustering rely exclusively on the metrics of the embedding space, and for latent variables of the encoder-decoder network, it cannot be expected to be Euclidean. Conversely, GMM and density-based clustering methods (e.g. DBSCAN[71]) rely on the local metrics, in this case similarity of physical mechanisms, and hence are expected to provide better results. The color coding in Fig. 6 (a) corresponds to GMM separation of the latent variables into 8 components. Once the GMM centroids in the latent space are established, the decoder part of the *im2spec* network can be used to reconstruct characteristic hysteresis loop behaviors. The result in Fig. 6 (d) shows clear variability of behaviors from well-developed hysteresis loops to strongly constricted ones. Note that similar analysis can be performed for any dimensionality of latent space and generally demonstrates consistent behavior.

By extending this framework, we can obtain insight into which domain structures give rise to the characteristic hysteresis loops illustrated in Fig. 6 (e). While the translation from domain images into the latent variables is performed in the encoder part of network (and hence inverse process is performed only as a part of backpropagation training), here we utilize the fact that, for meaningful representation, the encoder interpolation is a continuous function. Hence, to establish the relevant behaviors we choose the 100 pixels closest to the centroid of the GMM unmixing and determined the average domain structure corresponding to these points. The comparison of the hysteresis loops and corresponding domain structure elements in Fig. 6 (d) and 6 (e) illustrates strong correspondence between the two.

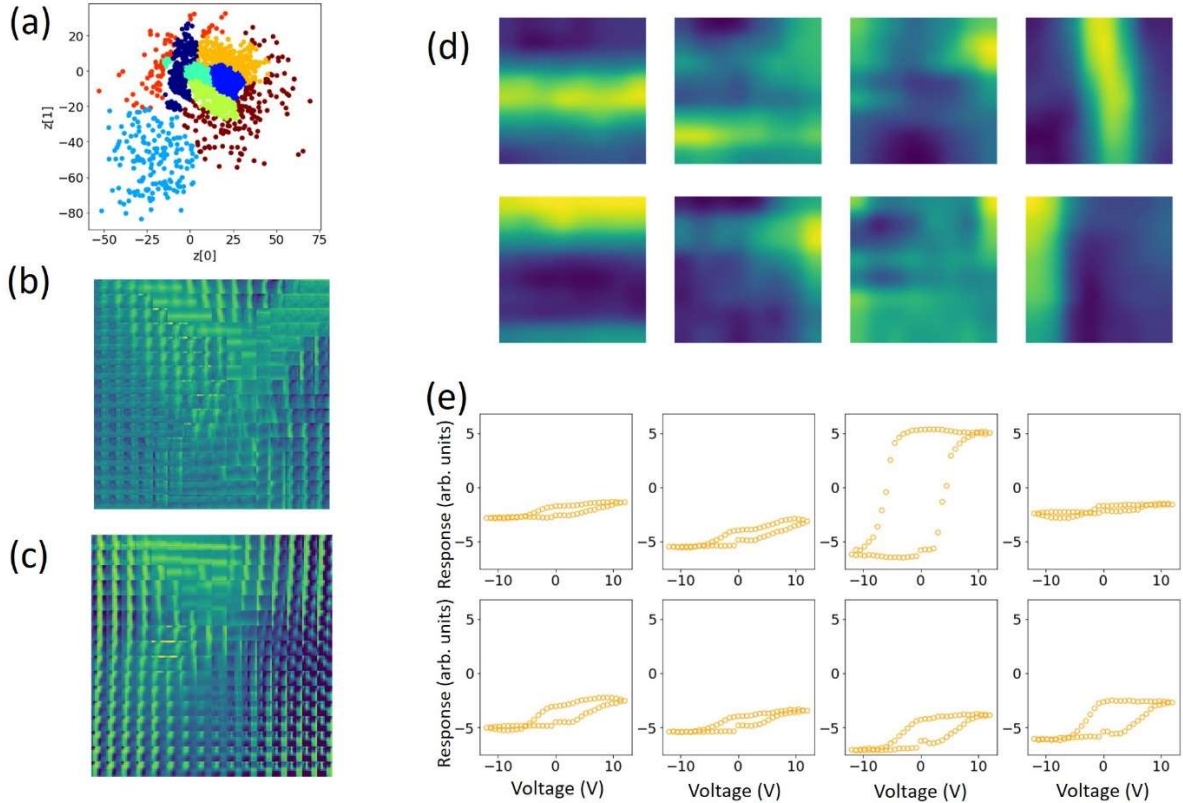

**Figure 7.** (a) Latent space of the *spec2im* model and (b) learned 2D manifold with different domain behaviors reconstructed from latent variables by the *spec2im* decoder. (c) "inverse" reconstruction of domain behaviors with *im2spec* model encoder. (d) The domain structure elements obtained from the reconstruction from the latent variables at the GMM centroids and (e) the corresponding hysteresis loops.

Similar analysis can be performed for the *spec2im* model. Here, feature and target sets are interchanged, thus allowing for prediction of images from spectral data (central and right columns in Fig. 4) and establishing a relationship between the spectra and local domain patterns. The structure of the latent space for 2 latent variables is shown in Fig. 7 (a) and the analysis of the encoder and decoder parts of the *spec2im* network are shown in Figure 7 (d, e). Here, again, the latent space was clustered using the GMM method, and the decoded images corresponding to cluster centroids are shown in Fig. 7 (d). The images show clear spatial structure corresponding to dissimilar domain boundary orientations and distances from the center points to the boundary. The corresponding hysteresis loops defined as averages of the hysteresis loops at 10 locations, which in the latent representation, are closest to the centroid, are shown in Figure 7 (e).

Finally, we comment upon the universality of this approach as applied to diverse data sets. We note that the application of deep learning in computer science typically rely on well-established data sets, and in biology and medicine on the large integrated data bases. As such, these areas typically deal with the in-distribution data, meaning the "new" data (either test data, or new results) are well within the variability of the training data set. At the same time, experimental measurements in physical sciences often deal with the relatively small data sets with strong observational biases (e.g. microscope calibration), and hence generally expecting the transferability between individual experiments is reasonable only if measurements are perfectly calibrated and reproducible, and a sufficiently large data set is available. Here, we have shown that a model trained on one section of the image will generalizes towards the other (previously unseen/unused) section of the image, which is roughly equivalent to applying the model trained on data from one scan area (sample region) to make a prediction on data from a different scan area (sample region) acquired with the same parameters. At the same time, applying the model trained on a single dataset to data acquired from a different experiment will not work unless: the scale, orientation and probe condition are close to the original training data; the microscope is calibrated identically; and the surface state of material is the same. These issues can be addressed by creating a much larger, ImageNet-like, datasets for different materials which also accounts for different experimental conditions (e.g., different state of the probe). Similarly, this necessitates rapid adoption of the PFM calibration methods in the standard imaging workflows. This, however, will likely require a community effort and is beyond the scope of the current work.

**Conclusion**

To summarize, here we introduce the convolutional encoder-decoder based machine learning workflow to explore the correlation between the local structure and spectral responses, embodied as image to spectral (*im2spec*) and spectral to image (*spec2im*) networks. This approach is used to establish the correlation between local domain structures and switching behavior in ferroelectric materials, and allows for the identification of associated domain structure elements. The analysis holds for the data sets acquired under individual conditions, but as expected requires matching of sample orientation, surface state of the material, and microscope parameters. The regions with high prediction error define obvious targets for in-depth studies, either human based or in automated experiment.

This approach provides the universal workflow to establish the presence of universal correlation between local spectral responses and adjacent structure, and can be universally applied to spectral imaging techniques such as PFM, scanning tunneling microscopy (STM) and spectroscopy, and electron energy loss spectroscopy (EELS) in scanning transmission electron microscopy (STEM). The code is available as an executable Jupyter notebook at https://git.io/JLs7H

We further note that the encoder-decoder approach proposed here represents the simplest case of the possible network architecture, and sets a clear case for the incorporation of the network architectures invariant towards rotations and scaling,[72-74] as well as incorporation of physics-based priors. This will allow generalizing a neural network performance to other sample orientations (insofar it does not affect image formation mechanisms, e.g. coupling between normal and longitudinal responses in PFM). This further highlights the increasing role of quantitativeness, i.e. capability of the imaging method to provide materials-specific behaviors independent from imaging system, in physical imaging.

**Acknowledgements:**

This research was conducted at the Center for Nanophase Materials Sciences, which also provided support (RKV, MZ, SKV) and is a US DOE Office of Science User Facility. The PFM experiment was supported by the U.S. Department of Energy, Office of Science, Basic Energy Sciences, Materials Sciences and Engineering Division (KK). The authors are thankful to Prof. Hiroshi Funakubo (Tokyo Institute of Technology) for providing PbTiO$_3$ thin film samples. The authors would also like to express their gratitude to two of the reviewers, whose comments on the proposed methodology resulted in significant improvement of model's evaluation.

**TOC**

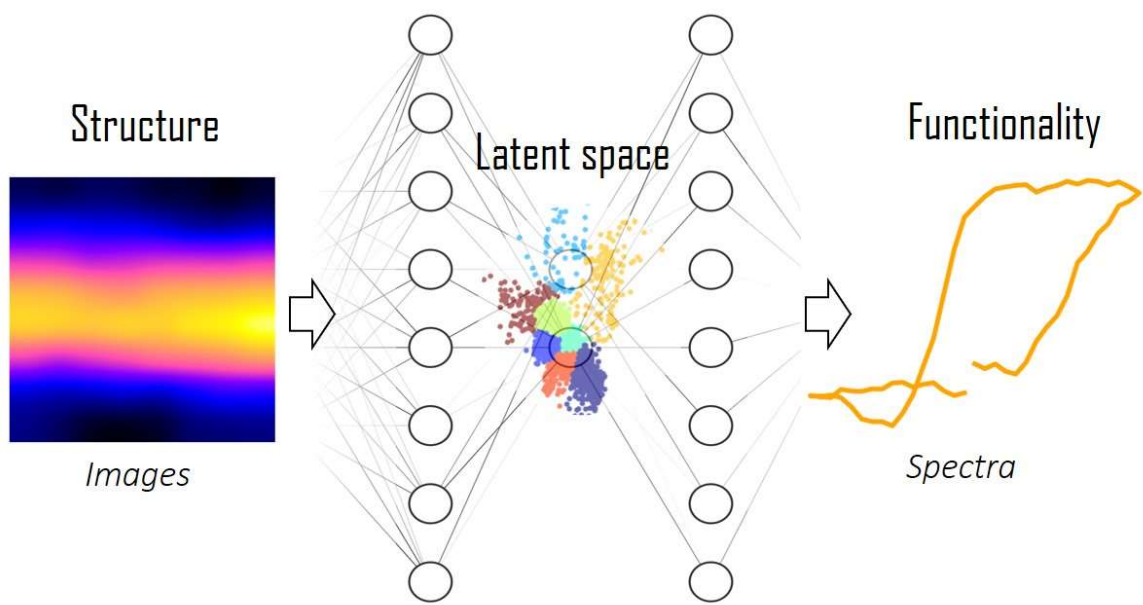